%% file: paper.tex
\documentclass{PoS}
\usepackage{amsmath,amssymb,amsfonts}
\usepackage{slashed}
\usepackage{bbm}
\usepackage{cite}
\usepackage[utf8]{inputenc}
\usepackage[english]{babel}

\title{$G_2$-QCD: Spectroscopy and the phase diagram at zero temperature and finite density}
\ShortTitle{$G_2$-QCD: Spectroscopy}
\author{\speaker{Bjoern H. Wellegehausen}\\%
        Justus-Liebig-University Giessen and Friedrich-Schiller-University Jena\\
        E-mail: \email{bjoern.wellegehausen@theo.physik.uni-giessen.de}}
\author{Lorenz von Smekal\\%
        Justus-Liebig-University Giessen and TU Darmstadt\\
        E-mail: \email{lorenz.smekal@physik.tu-darmstadt.de}}
\author{Axel Maas and Andreas Wipf\\
	Friedrich-Schiller-University Jena\\
	E-mail: \email{axelmaas@web.de}, \email{wipf@tpi.uni-jena.de}}

\abstract{
Due to the fermion sign problem, standard lattice Monte-Carlo method for QCD fail at small temperatures and high baryon densities.
$G_2$-QCD, QCD with the gauge group $SU(3)$ replaced by the exceptional Lie group $G_2$, can be simulated using lattice techniques at these densities, 
and can therefore provide an illustration of the possible phase structure. 
Here we present a systematic investigation of the ground-state hadronic spectrum using lattice simulations for different quark masses in several hadronic sectors. 
We then show that the different hadronic scales of Goldstone bosons, intermediate bosons, and baryons is reflected in the phase structure at finite density.
}

\FullConference{31st International Symposium on Lattice Field Theory LATTICE 2013\\
		 July 29 - August 3, 2013\\
		 Mainz, Germany}
\usepackage{etex}
\usepackage{graphicx}
\usepackage{bbm,bm}
\usepackage{mathtools}
\usepackage{color,array,tabularx}
\usepackage{stackrel}
\usepackage[small,bf]{caption}
\usepackage{pst-all}
\usepackage{psfrag}
\usepackage{pstricks,pst-grad,color,shadow}
\usepackage{tikz}
\usetikzlibrary{shapes,arrows,backgrounds}
\usepackage{soul}
\usepackage{placeins}
\usepackage{latexsym}
\usepackage{keyval}
\usepackage{simplewick}
\usepackage{epigraph}
\usepackage{amsthm}

\definecolor{darkgreen}{rgb}{0,0.73,0}

\newcommand{\erw}[1]{\left \langle #1 \right \rangle}

\newcommand{\algebra}[1]{\mathfrak{#1}}

\DeclareMathOperator{\tr}{tr}

\newcommand{\ii}{\mathrm{i}}

\newcommand{\chargec}{\mathsf{C}}

\newcommand{\egamma}{{\gamma_\mathsf{E}}}

\graphicspath{{./}{./plots/}{./publPlots/}}

\begin{document}

\section{Introduction}

\noindent QCD at finite baryon density suffers from a severe sign problem and its phase diagram can therefore not be treated with 
standard Monte-Carlo methods \cite{Gattringer:2010zz}. 
Several approaches to QCD have been investigated to circumvent this problem, but so far all these approaches failed to describe the 
phase diagram at high baryon density and small temperature.
A different strategy is to investigate QCD-like theories, as for instance
2-color QCD \cite{Kogut:2000ek,Hands:2011ye}, to gain insight into gauge theories at finite density. 
However, these theories should share as much features as possible with QCD. Here we replace the 
gauge group $SU(3)$ by the exceptional Lie group $G_2$. $G_2$-QCD is a gauge theory with fermionic baryons and 
fundamental quarks \cite{Holland:2003jy,Maas:2012wr} and it can be simulated without sign problem at finite density and temperature.
In the present article we discuss the theoretical foundations of (lattice) spectroscopy for $G_2$-QCD  and present our results 
for the vacuum spectrum of states and the phase diagram at zero temperature obtained from lattice simulations on a rather 
small $8^3 \times 16$ lattice. We find that we can observe structures at finite density at scales which correspond to scales 
we find in the hadronic spectrum. Especially, we find onsets of transitions at scales corresponding to the Goldstone scale, 
the intermediate boson scale, and the baryonic scale. The results indeed suggest that the theory has a rich phase structure, 
and that baryonic-dominated regions of the phase diagram may exist.

\section{Chiral symmemtry and baryon number in $G_2$-QCD}\label{sg2props}

\noindent The Euclidean action of $N_\text{f}$ flavour $G_2$-QCD with baryon chemical potential $\mu$ reads
\begin{equation}
\begin{aligned}
S=&\int d^4 x \tr \left \lbrace-\frac{1}{4} F_{\mu\nu}F^{\mu\nu} + \sum \limits_{n=1}^{N_\text{f}} \bar{\Psi}_n \, D[A,m,\mu] \, \Psi_n\right \rbrace \quad \text{with} \\ 
D[A,m,\mu]=&\egamma^\mu (\partial_\mu-g A_\mu)-m +\egamma_0 \mu,
\label{eqn:actionQCD}
\end{aligned}
\end{equation}
where the gauge group is the exceptional Lie group $G_2$. The fundamental representations of $G_2$ are $7$-dimensional and 
$14$-dimensional, the latter coinciding with the adjoint representation. The elements of $G_2$ can be viewed as
elements of $SO(7)$ subject to seven independent cubic constraints for the
$7$-dimensional matrices $g$ representing $SO(7)$
\cite{Holland:2003jy},
\begin{equation}
\label{eq:g2constraint}
T_{abc} = T_{def}\,g_{da}\,g_{eb}\,g_{fc},
\end{equation}
where $T$ is a total antisymmetric tensor. Since $G_2$ is a subgroup of $SO(7)$, all representations are real. 
The Dirac operator satisfies
\begin{equation}
D(\mu)^\dagger \,\gamma_5=\gamma_5\,D(-\mu^*) \quad \text{and} \quad D(\mu)^* \,T=T\,D(\mu^*)
\end{equation}
with $T=C\gamma_5$, $ T^*\, T=-\mathbbm{1}$, $T^\dagger=T^{-1}$ and charge conjugation matrix $C$. If such a unitary operator $T$ exists then the
eigenvalues of the Dirac operator come in complex conjugate pairs, all real
eigenvalues are doubly degenerate 
\cite{Kogut:2000ek,Hands:2000ei} and thus
\begin{equation}
\det D[\,A,m,\mu]\geq 0 \quad \text{for} \quad \mu \in \mathbbm{R}.
\end{equation}
Since the gauge field $A_\mu$ is real and anti-symmetric, it is possible to write the matter part of the action \eqref{eqn:actionQCD} for $\mu=0$ as a sum over
$2N_\mathsf{f}$ Majorana spinors $\lambda_n$
\begin{equation}
\mathcal{L}[\Psi,A]=\bar{\Psi}\, D[A,m,0] \,\Psi
=\bar{\lambda}\, D[A,m,0] \,\lambda
\label{eqn:actionQCDMaj}
\end{equation}
with $\lambda=(\chi\,,\eta)=(\lambda_1,\dots,\lambda_{2N_\text{f}})$. It follows that $G_2$-QCD possesses an extended flavour symmetry \cite{Kogut:2000ek} compared to QCD. 
The action is invariant under the $SO(2N_\mathsf{f})_\mathsf{V}$ vector transformations
$\lambda \mapsto  e^{\beta \otimes \mathbbm{1}}\lambda$
with a real and antisymmetric $\beta\in\algebra{so}(2 N_\mathsf{f})$ 
and the axial transformations
 $\lambda \mapsto e^{\ii \, \alpha \otimes \gamma_5}\lambda$
with a real symmetric matrix $\alpha$.
Due to the Majorana constraint left- and right-handed spinors cannot be rotated independently
and the general transformation is a composition of axial- and vector transformations, leading to a $U(2N_\mathsf{f})$ symmetry group, in agreement with the results in \cite{Holland:2003jy}.
Following the same arguments as in QCD it is expected that 
the axial $U(1)$ is broken by the axial anomaly such 
that only a $SU(2N_\mathsf{f})\times \mathbbm{Z}(2)_\mathsf{B}$
chiral symmetry remains. In the presence of a non-vanishing Dirac mass term (or a non-vanishing chiral
condensate) the theory is no longer invariant under the axial
transformations. Therefore the non-anomalous chiral symmetry is
expected to be broken explicitly (or spontaneously) to its
maximal vector subgroup,
\begin{equation}
SU({2N_\mathsf{f}})\otimes
\mathbbm{Z}(2)_\mathsf{B} \mapsto 
SO(2 N_\mathsf{f})_\mathsf{V}\otimes \mathbbm{Z}(2)_\mathsf{B},
\end{equation}
Since baryon chemical potential is an off-diagonal term in Majorana flavour space, the remaining chiral symmetry at finite baryon chemical potential is the same as in QCD,
\begin{equation}
SU({2N_\mathsf{f}})\otimes \mathbbm{Z}(2)_\mathsf{B} \mapsto 
SU(N_\mathsf{f})_\mathsf{A}\otimes SU(N_\mathsf{f})_\mathsf{V} \otimes U(1)_\mathsf{B}/\mathbbm{Z}(N_\mathsf{f}).
\end{equation}
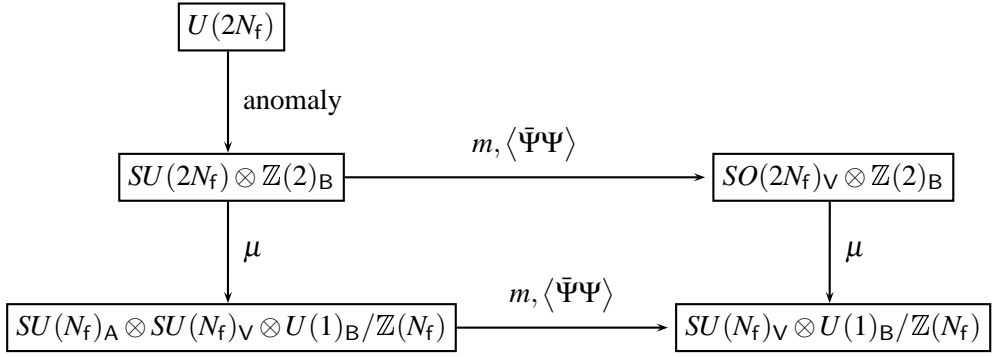
\begin{figure*}[tb]
\psset{xunit=1cm,yunit=1cm,runit=1cm}
\begin{center}
\begin{pspicture}(0,0.5)(12,4.5)
\rput(2,4){\rnode{A}{
\psframebox{$U(2N_\mathsf{f})$}}}
\rput(2,2){\rnode{B}{
\psframebox{$SU({2N_\mathsf{f}})
\otimes \mathbbm{Z}(2)_\mathsf{B}$}}}
\rput(10,2){\rnode{C}{
\psframebox{$SO(2 N_\mathsf{f})_\mathsf{V}\otimes \mathbbm{Z}(2)_\mathsf{B}$}}}
\rput(2,0){\rnode{D}{
\psframebox{$SU(N_\mathsf{f})_\mathsf{A}\otimes
SU(N_\mathsf{f})_\mathsf{V} \otimes
U(1)_\mathsf{B}/\mathbbm{Z}(N_\mathsf{f})$}}} 
\rput(10,0){\rnode{E}{
\psframebox{$SU(N_\mathsf{f})_\mathsf{V} \otimes
U(1)_\mathsf{B}/\mathbbm{Z}(N_\mathsf{f})$}}}
\ncline{->}{A}{B}\naput{anomaly}
\ncline{->}{B}{C}\naput{$m, \erw{\bar{\Psi} \Psi}$}
\ncline{->}{C}{E}\naput{$\mu$}
\ncline{->}{D}{E}\naput{$m,\erw{\bar{\Psi} \Psi}$}
\ncline{->}{B}{D}\naput{$\mu$}
\end{pspicture}
\end{center}
\vskip5mm
\caption{Pattern of chiral symmetry breaking in $G_2$-QCD.}
\label{fig:chiralSymmG2QCD}
\end{figure*}
The final pattern of chiral symmetry breaking of $G_2$-QCD is shown in figure \ref{fig:chiralSymmG2QCD}. If chiral symmetry is spontaneously broken, the axial chiral multiplet becomes
massless, according to the Goldstone theorem. In contrast to QCD, because of the enlarged chiral symmetry group, already in the case of a single Dirac flavour a
non-trivial chiral symmetry is present, and chiral symmetry breaking can be
observed. In this $N_\mathsf{f}=1$ case chiral symmetry is given by $SU(2) \otimes \mathbbm{Z}(2)_\mathsf{B}$
leading to two Goldstone bosons
\begin{equation}
\begin{aligned}
d(0^{++}) = \bar{\Psi}^\chargec \gamma_5 \Psi-\bar{\Psi}\gamma_5\Psi^\chargec \quad \text{and} \quad
d(0^{+-}) = \bar{\Psi}^\chargec\gamma_5\Psi+\bar{\Psi}\gamma_5\Psi^\chargec.
\end{aligned}
\end{equation}
with baryon number $n_\text{B}=2$\footnote{In our definition baryon number counts the difference of quarks and anti-quarks}. They are scalar diquarks instead of
pseudoscalar mesons as in QCD. 

\section{Spectroscopy for $N_f=1$ $G_2$-QCD}\label{sspectheo}

\noindent Since the centre of $G_2$ is trivial, we expect to find bound states with every integer baryon number. Beside the states that are also present in QCD, Mesons with baryon number $n_\text{B}=0$ and nucleons with $n_\text{B}=3$,  there are additional states like diquarks with $n_\text{B}=2$ or more exotic bound states of gluons and quarks, for example a hybrid with $n_\text{B}=1$. In the following we give an overview over our implementation of possible bound states for $N_f=2$ $G_2$-QCD, see tables \ref{tab:nb02} and \ref{tab:nb13}.
\begin{table}[htb]
\begin{center}
\begin{tabular}{|c|c|c|c|c|c|}
\hline Name & $\mathcal{O}$  & J & P & C\\
\hline $\pi$ & $\bar{u} \gamma_5 d$  & 0 & - & + \\
\hline $\eta$ & $\bar{u} \gamma_5 u$  & 0 & - & + \\
\hline $a$ & $\bar{u}d$  & 0 & + & + \\
\hline $f$ & $\bar{u}u$  & 0 & + & + \\
\hline $\rho$ & $\bar{u}\gamma_\mu d$  & 1 & - & + \\
\hline $\omega$ & $\bar{u}\gamma_\mu u$  & 1 & - & + \\
\hline $b$ & $\bar{u}\gamma_5 \gamma_\mu d$  & 1 & + & + \\
\hline $h$ & $\bar{u}\gamma_5 \gamma_\mu u$ & 1 & + & + \\
\hline
\end{tabular}\hskip5mm
\begin{tabular}{|c|c|c|c|c|}
\hline Name & $\mathcal{O}$ & J & P & C\\
\hline $d(0^{++})$ & $\bar{u}^\chargec \gamma_5 u + c.c.$  & 0 & + & + \\
\hline $d(0^{+-})$ & $\bar{u}^\chargec \gamma_5 u - c.c.$  & 0 & + & - \\
\hline $d(0^{-+})$ & $\bar{u}^\chargec u + c.c.$  & 0 & - & + \\
\hline $d(0^{--})$ & $\bar{u}^\chargec u - c.c.$  & 0 & - & - \\
\hline $d(1^{++})$ & $\bar{u}^\chargec \gamma_\mu d - \bar{d}^\chargec \gamma_\mu u + c.c.$  & 1 & + & + \\
\hline $d(1^{+-})$ & $\bar{u}^\chargec \gamma_\mu d - \bar{d}^\chargec \gamma_\mu u - c.c.$  & 1 & + & - \\
\hline $d(1^{-+})$ & $\bar{u}^\chargec \gamma_5 \gamma_\mu d - \bar{d}^\chargec \gamma_5 \gamma_\mu u + c.c.$  & 1 & - & + \\
\hline $d(1^{--})$ & $\bar{u}^\chargec \gamma_5 \gamma_\mu d - \bar{d}^\chargec \gamma_5 \gamma_\mu u - c.c.$  & 1 & - & - \\
\hline
\end{tabular}\\
\end{center}
\caption{Bosonic bound states with baryon number $n_\text{B}=0$ (left table) and $n_\text{B}=2$ (right table).}
\label{tab:nb02}
\end{table}
\begin{table}[htb]
\begin{center}
\begin{tabular}{|c|c|c|c|c|}
\hline Name & $\mathcal{O}$ & J & P & C\\
\hline $N'$ & $T^{abc}(\bar{u}_{a} \gamma_5 d_{b}) u_{c}$  & 1/2 & $\pm$ & $\pm$ \\
\hline $\Delta'$ & $T^{abc} (\bar{u}_{a} \gamma_\mu u_{b}) u_{c}$  & 3/2 & $\pm$ & $\pm$ \\
\hline Hybrid & $\epsilon_{abcdefg} u^a F_{\mu\nu}^{bc} F_{\mu\nu}^{de} F_{\mu\nu}^{fg}$  & 1/2 & $\pm$ & $\pm$ \\
\hline
\end{tabular}\hskip5mm
\begin{tabular}{|c|c|c|c|c|}
\hline Name & $\mathcal{O}$ & J & P & C\\
\hline $N$ & $T^{abc}(\bar{u}_{a}^\chargec \gamma_5 d_{b}) u_{c}$  & 1/2 & $\pm$ & $\pm$ \\
\hline $\Delta$ & $T^{abc} (\bar{u}_{a}^\chargec \gamma_\mu u_{b}) u_{c}$  & 3/2 & $\pm$ & $\pm$ \\
\hline
\end{tabular}\\
\end{center}
\caption{Fermionic bound states with baryon number $n_\text{B}=1$ (left table) and $n_\text{B}=3$ (right table).}
\label{tab:nb13}
\end{table}
In all tables $\mathcal{O}$ is the interpolating operator used to extract the mass in the lattice simulation and $J$, $P$, $C$  the spin, parity and charge conjugation quantum numbers.
In our lattice simulations the $N_f=2$ $G_2$-QCD states are included by partial quenching, i.\ e.\ two valence quark flavor, but only one sea quark flavor. 
If isospin is unbroken, the masses of flavour singlet diquarks and flavour non-singlet mesons are degenerate. 
For example the diquark correlation function
\begin{equation}
C_{d}(x,y)=\erw{ \contraction[2ex]{}{\bar{\chi}(x)}{\gamma_5
\chi(x)
\bar{\chi}(y) \gamma_5}{\chi(y)} 
\contraction[1ex]{\bar{\chi}(x) \gamma_5 }{\chi(x)}{}{\bar{\chi}(y)}
\bar{\chi}(x)\gamma_5 \chi(x)\, \bar{\chi}(y)\gamma_5,
\chi(y)}
\end{equation}
contains only connected contributions. The corresponding correlation function for the $\eta$ meson reads
\begin{equation}
C_{\eta}(x,y)=\erw{{\eta}(x)\, {\eta}^\dagger(y)}
=2\erw{
\contraction[1ex]{}{\bar{\chi}(x)}{\gamma_5}{\chi(x)} 
\contraction[1ex]{\bar{\chi}(x) \gamma_5
\chi(x)}{\bar{\chi}(y)}{\gamma_5}{\chi(y))} \bar{\chi}(x)\gamma_5 \chi(x)\,
\bar{\chi}(y)\gamma_5 \chi(y)}+C_{d}(x,y)\label{etacorr}
\end{equation}
where the difference to the diquark correlation function is only the disconnected contribution, showing that the $\pi$ has the same mass as the $d(0^{+})$. Analog relations lead to
\begin{equation}
 m_{d(0^+)}=m_{\pi(0^-)}, \quad
m_{d(0^-)}=m_{a(0^+)}, \quad
m_{d(1^+)}=m_{\rho(1^-)}, \quad
m_{d(1^-)}=m_{b(1^+)}.
\end{equation} 
Thus, for every diquark there is a flavour non-singlet meson with the same mass but opposite parity.

\section{Lattice spectroscopy results}\label{sspectrum}

\begin{figure}[htb]
\begin{center}
\scalebox{0.55}{\input{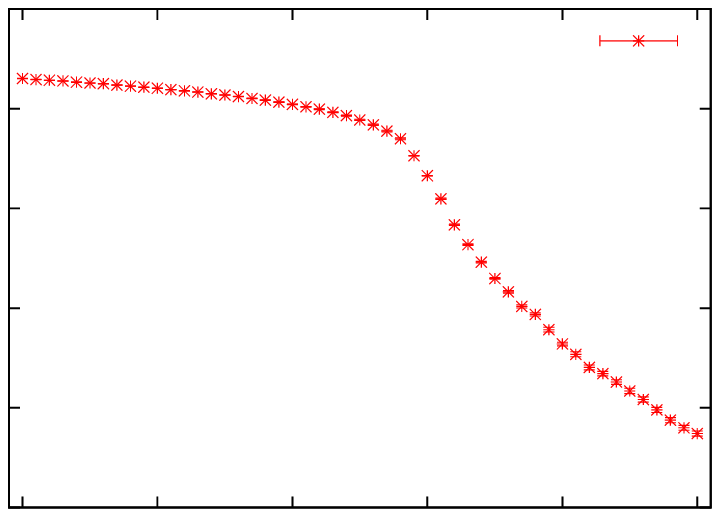}}\hskip3mm
\scalebox{0.55}{\input{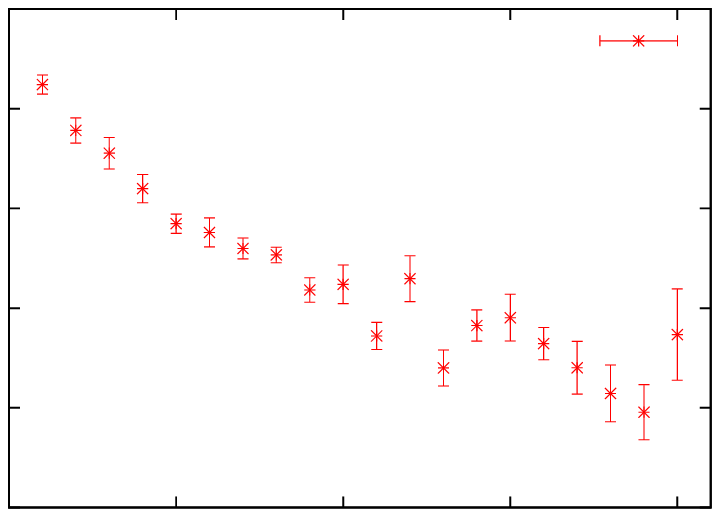}}\hskip3mm
\scalebox{0.55}{\input{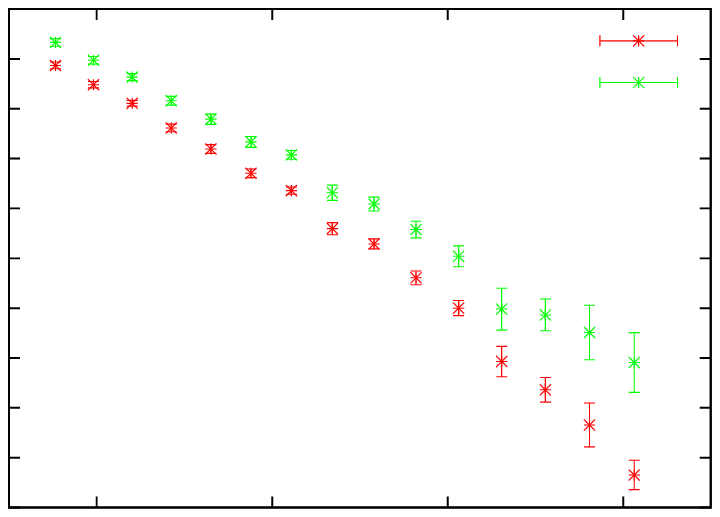}}
\end{center}
\caption{\textbf{Left panel:} Mass of the pseudo Goldstone boson as a function of $\beta$ for $\kappa=0.147$. \textbf{Center panel:} Mass of the proton as a function of $\beta$ for $\kappa=0.147$. \textbf{Right panel:} Mass of the $0^{+}$ and the $1^{+}$ diquark as a function of $\kappa$ for $\beta=0.96$.}
\label{fig:massDiquarkProton}
\end{figure}

\noindent
In order to fix our parameters we compute the scalar and vector diquark masses and the proton mass for different parameters
of the inverse gauge coupling $\beta$ and the hopping parameter $\kappa$ on a $8^3 \times 16$ lattice. The results are shown in figure \ref{fig:massDiquarkProton}.
The mass ratios between the scalar and vector diquarks and between the scalar diquarks and nucleons allow us to estimate the distance from the chiral limit, since they should go to zero in the chiral limit while they approach $1$ and $2/3$ for heavy quarks.
\begin{table*}[htb]
\begin{tabular}{|c|c|c|c|c|c|c|c|c|}
\hline  Ensemble & $\beta$ & $\kappa$ & $m_{d(0^+)} a$ & $m_N a$  & $m_{d(0^+)}$ [MeV] & $a$ [fm] & $a^{-1}$ [MeV]& MC\\
\hline Heavy & $1.05$ & $0.147$ & $0.59(2)$ & $1.70(9)$ & $326$ & 0.357(33) & 552(50) & 7K   \\
\hline Light & $0.96$ & $0.159$ & $0.43(2)$ & $1.63(13)$ & $247$ & 0.343(45) & 575(75) & 5K \\
\hline
\end{tabular}\\
\caption{Parameters for two different ensembles. All results are from a $8^3\times 16$ lattice.}
\label{tab:ensembles}
\end{table*}
In the following we discuss two different ensembles with parameters shown in table \ref{tab:ensembles}. Our mass scale is set by the proton mass, $m_N=938$ MeV.
\begin{figure}[htb]
\begin{center}
\scalebox{0.7}{\input{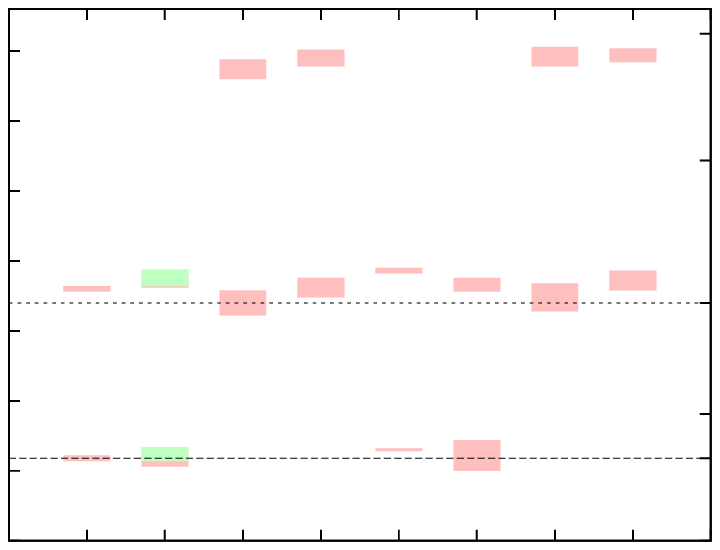}}\hskip10mm
\scalebox{0.7}{\input{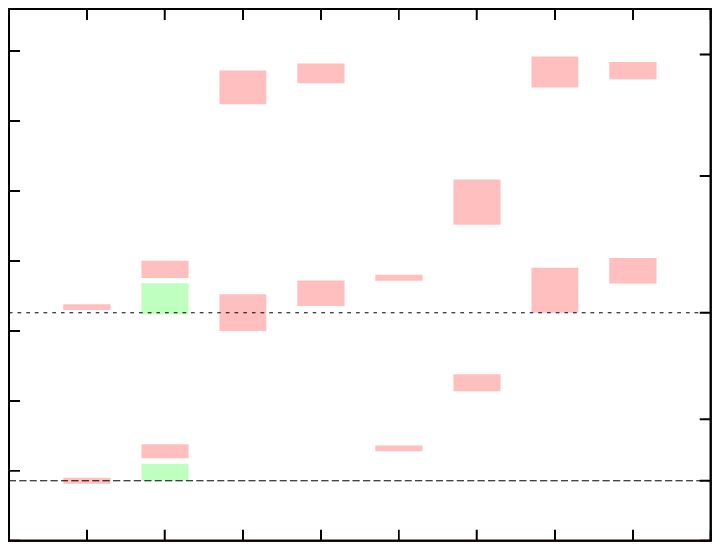}}
\end{center}
\caption{The mass spectrum of the heavy (left panel) and light (right panel) ensemble is shown.}
\label{fig:massSpectrum}
\end{figure}
For the heavy quark ensemble (Fig.~\ref{fig:massSpectrum}, left panel) the diquark masses are almost degenerate and the $\eta$ has essentially the same mass as the diquarks. 
For the nucleons there is almost no mass splitting between parity even and odd states. 
In the light ensemble (Fig.~\ref{fig:massSpectrum}, right panel), the diquark masses are no longer degenerate. 
We observe a significant mass splitting between parity even and odd states as well as between scalar and vector diquarks. 
Especially, the Goldstone boson becomes the lightest state, with the $\eta$ also being somewhat heavier. 
For the nucleons we also observe different masses for parity even and odd states and the spin 1/2 and spin 3/2 representations. 
We find three clearly different scales in the light spectrum: A Goldstone scale, an intermediate boson scale set by the remaining diquarks, 
and the nucleon scale set by the $N$ and $\Delta$.

\section{$G_2$-QCD at zero temperature and finite baryon density}\label{sdensity}

\begin{figure}[htb]
\begin{center}
\scalebox{0.6}{\input{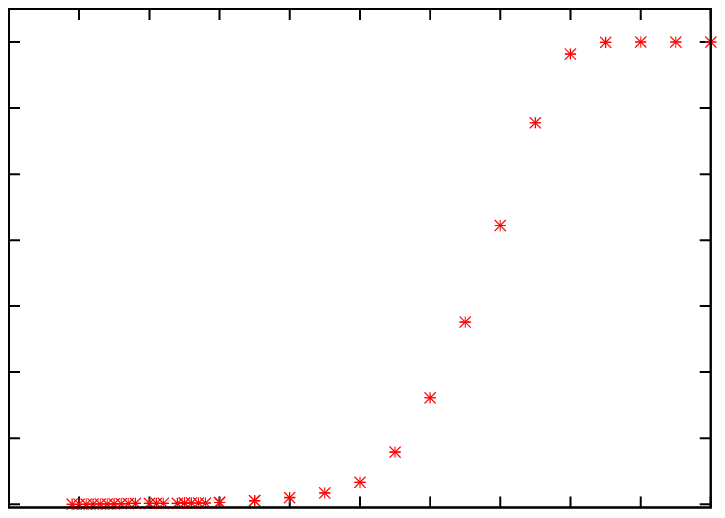}}\hskip10mm
\scalebox{0.6}{\input{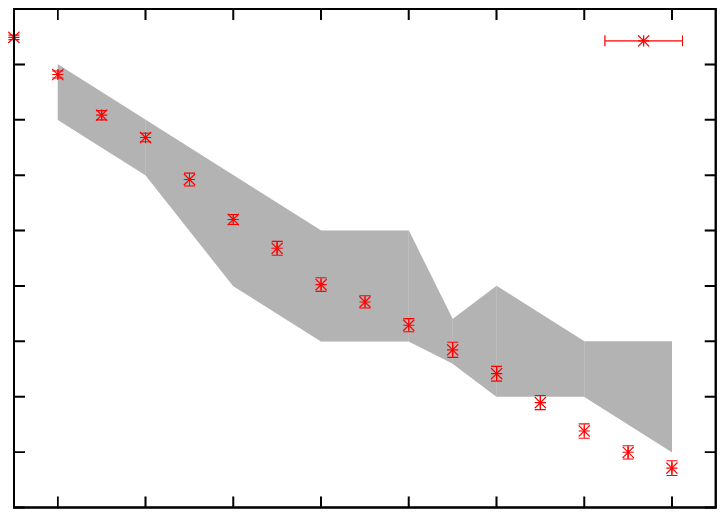}}
\end{center}
\caption{\textbf{Left panel:} The quark number density as a function of chemical potential are shown. 
\textbf{Right panel:} The onset transition observed in the quark number density is compared to half of the mass of the lightest state, 
the $0^+$ diquark, for different gauge couplings $\beta$.}
\label{fig:saturation}
\end{figure}

\noindent
We will now show that the different hadronic scales observed in the spectra in Fig.~\ref{fig:massSpectrum} reflect themselves in the structure of the finite density phase diagram.
In figure \ref{fig:saturation} (left panel) we show the quark number density $n_q$ from vanishing chemical potential up to saturation.
We observe that for small values of the chemical potential the system
remains in the vacuum, i.\ e.\ the quark number density vanishes, which is expected
due to the silver blaze property. When increasing the chemical
potential further the quark number density starts rising, indicating that baryonic matter is
present and the system is no longer in the vacuum state. At even larger values of $\mu$ the quark number density
saturates, in agreement with the theoretical prediction of $n_{q,\text{max}}=2 N_\mathsf{c}=14$ \cite{Maas:2012wr}.
A closer look into this phase diagram at zero temperature, see Fig.~\ref{fig:saturation} (right panel), shows the manifestation of the silver blaze property for baryon chemical potential: The quark number density shows an onset transition to a non-vacuum state when chemical potential reaches half of the mass of the lightest baryon, the Goldstone $0^+$ diquark.
\begin{figure}[htb]
\begin{center}
\scalebox{0.7}{\input{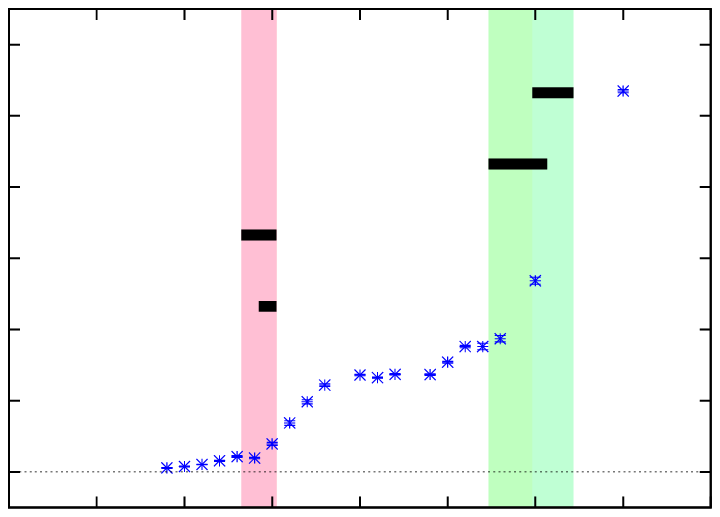}}\hskip10mm
\scalebox{0.7}{\input{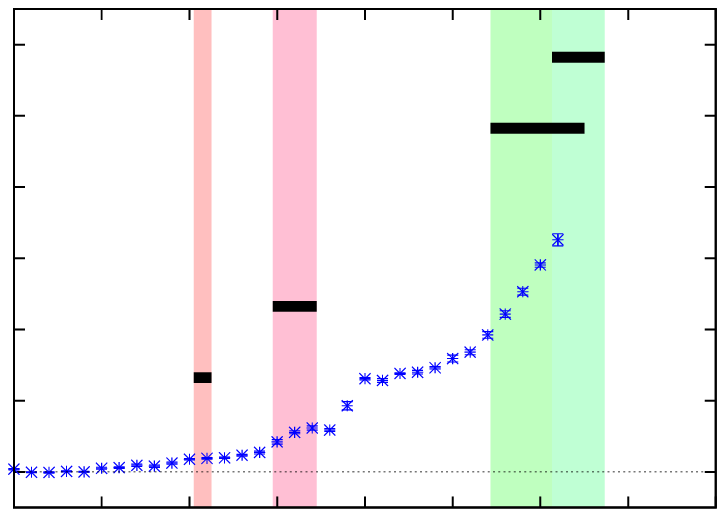}}
\end{center}
\caption{Shown is the quark number density compared to baryon mass divided by baryon number for the heavy ensemble (left panel) and the light ensemble (right panel).}
\label{fig:onsets}
\end{figure}
For larger values of chemical potential a series of plateaus develops where the quark
number density is almost constant, see figure \ref{fig:onsets}. At around $\mu=0.6$ for the heavy ensemble and $\mu=0.55$ for the light ensemble the quark number density starts increasing again and no further plateau is observed. 
For the heavy ensemble, in addition to the silver blaze transition due to the diquark states we find good agreement of the $\Delta$ mass with the point where the quark number density increases without building a plateau.
For the light ensemble the two transitions at $\mu\approx 0.22$ and $\mu \approx 0.32$, each followed by a plateau, can be related to the observation of the splitting of the $0^+$ and $0^-$ diquark masses. Again the transition at $\mu \approx 0.55$ is in good agreement with the $\Delta$ mass divided by three.
For both ensembles our observation is thus that transitions in the quark number density coincide with hadron masses divided by their baryon number. For a bosonic hadron a plateau is formed after the transition while for a fermionic hadron the quark number density increases further with increasing chemical potential. In both ensembles we observe also a transition at $\mu\approx0.52$ (heavy ensemble) and $\mu\approx0.38$ (light ensemble) that does not coincide with any of our spectroscopic states.  Since this transition is followed by a plateau we speculate that this state might be a bosonic hadron. A possible candidate could for example be a bound state of four quarks.

\section{Conclusions}\label{sconclusions}

\noindent On a rather small lattice we have presented the hadronic spectrum of $G_2$-QCD obtained with lattice Monte-Carlo simulations. For sufficiently small quark masses different hadronic scales related to the Goldstone sector and nucleonic sector develop, quite similar to the situation in ordinary QCD. We have also shown that this scale hierarchy of the vacuum reflects itself in the phase structure at finite densities. The phase diagram at zero temperature shows a number of transitions which correlate with the scales of the hadron spectrum. This already indicates a very rich phase structure of the theory and ongoing investigations aim at a deeper understanding of the
phase diagram of $G_2$-QCD at finite density.

\begin{acknowledgments}
\noindent 
We are grateful to Jonivar Skullerud for helpful discussions. This work was supported by the Helmholtz International Center for FAIR within the LOEWE initiative of the State of Hesse. A.\ M.\ was supported by the DFG under grant number MA3935/5-1, B.\ W.\ by the DFG graduate school GRK 1523/1 and L.v.S. by the European Commission, FP-7-PEOPLE-2009-RG, No. 249203. Simulations were performed on the LOEWE-CSC in Frankfurt and the HPC cluster in Jena.
\end{acknowledgments}

\end{document}

%% file: plots/Mass_8_16.tex
\begingroup
\scriptsize
  \makeatletter
  \providecommand\color[2][]{%
    \GenericError{(gnuplot) \space\space\space\@spaces}{%
      Package color not loaded in conjunction with
      terminal option `colourtext'%
    }{See the gnuplot documentation for explanation.%
    }{Either use 'blacktext' in gnuplot or load the package
      color.sty in LaTeX.}%
    \renewcommand\color[2][]{}%
  }%
  \providecommand\includegraphics[2][]{%
    \GenericError{(gnuplot) \space\space\space\@spaces}{%
      Package graphicx or graphics not loaded%
    }{See the gnuplot documentation for explanation.%
    }{The gnuplot epslatex terminal needs graphicx.sty or graphics.sty.}%
    \renewcommand\includegraphics[2][]{}%
  }%
  \providecommand\rotatebox[2]{#2}%
  \@ifundefined{ifGPcolor}{%
    \newif\ifGPcolor
    \GPcolortrue
  }{}%
  \@ifundefined{ifGPblacktext}{%
    \newif\ifGPblacktext
    \GPblacktexttrue
  }{}%
  \let\gplgaddtomacro\g@addto@macro
  \gdef\gplbacktext{}%
  \gdef\gplfronttext{}%
  \makeatother
  \ifGPblacktext
    \def\colorrgb#1{}%
    \def\colorgray#1{}%
  \else
    \ifGPcolor
      \def\colorrgb#1{\color[rgb]{#1}}%
      \def\colorgray#1{\color[gray]{#1}}%
      \expandafter\def\csname LTw\endcsname{\color{white}}%
      \expandafter\def\csname LTb\endcsname{\color{black}}%
      \expandafter\def\csname LTa\endcsname{\color{black}}%
      \expandafter\def\csname LT0\endcsname{\color[rgb]{1,0,0}}%
      \expandafter\def\csname LT1\endcsname{\color[rgb]{0,1,0}}%
      \expandafter\def\csname LT2\endcsname{\color[rgb]{0,0,1}}%
      \expandafter\def\csname LT3\endcsname{\color[rgb]{1,0,1}}%
      \expandafter\def\csname LT4\endcsname{\color[rgb]{0,1,1}}%
      \expandafter\def\csname LT5\endcsname{\color[rgb]{1,1,0}}%
      \expandafter\def\csname LT6\endcsname{\color[rgb]{0,0,0}}%
      \expandafter\def\csname LT7\endcsname{\color[rgb]{1,0.3,0}}%
      \expandafter\def\csname LT8\endcsname{\color[rgb]{0.5,0.5,0.5}}%
    \else
      \def\colorrgb#1{\color{black}}%
      \def\colorgray#1{\color[gray]{#1}}%
      \expandafter\def\csname LTw\endcsname{\color{white}}%
      \expandafter\def\csname LTb\endcsname{\color{black}}%
      \expandafter\def\csname LTa\endcsname{\color{black}}%
      \expandafter\def\csname LT0\endcsname{\color{black}}%
      \expandafter\def\csname LT1\endcsname{\color{black}}%
      \expandafter\def\csname LT2\endcsname{\color{black}}%
      \expandafter\def\csname LT3\endcsname{\color{black}}%
      \expandafter\def\csname LT4\endcsname{\color{black}}%
      \expandafter\def\csname LT5\endcsname{\color{black}}%
      \expandafter\def\csname LT6\endcsname{\color{black}}%
      \expandafter\def\csname LT7\endcsname{\color{black}}%
      \expandafter\def\csname LT8\endcsname{\color{black}}%
    \fi
  \fi
  \setlength{\unitlength}{0.0500bp}%
  \begin{picture}(4534.00,3400.00)%
    \gplgaddtomacro\gplbacktext{%
      \csname LTb\endcsname%
      \put(384,512){\makebox(0,0)[r]{\strut{}0.0}}%
      \put(384,1086){\makebox(0,0)[r]{\strut{}0.5}}%
      \put(384,1660){\makebox(0,0)[r]{\strut{}1.0}}%
      \put(384,2235){\makebox(0,0)[r]{\strut{}1.5}}%
      \put(384,2809){\makebox(0,0)[r]{\strut{}2.0}}%
      \put(384,3383){\makebox(0,0)[r]{\strut{}2.5}}%
      \put(558,352){\makebox(0,0){\strut{}0.60}}%
      \put(1335,352){\makebox(0,0){\strut{}0.70}}%
      \put(2113,352){\makebox(0,0){\strut{}0.80}}%
      \put(2890,352){\makebox(0,0){\strut{}0.90}}%
      \put(3668,352){\makebox(0,0){\strut{}1.00}}%
      \put(4445,352){\makebox(0,0){\strut{}1.10}}%
      \put(2501,112){\makebox(0,0){\strut{}$\beta$}}%
    }%
    \gplgaddtomacro\gplfronttext{%
      \csname LTb\endcsname%
      \put(3788,3200){\makebox(0,0)[r]{\strut{}$m_{{d(0^+)}}$}}%
    }%
    \gplbacktext
    \put(0,0){\includegraphics{Mass_8_16}}%
    \gplfronttext
  \end{picture}%
\endgroup

%% file: plots/MassProton_8_16.tex
\begingroup
\scriptsize
  \makeatletter
  \providecommand\color[2][]{%
    \GenericError{(gnuplot) \space\space\space\@spaces}{%
      Package color not loaded in conjunction with
      terminal option `colourtext'%
    }{See the gnuplot documentation for explanation.%
    }{Either use 'blacktext' in gnuplot or load the package
      color.sty in LaTeX.}%
    \renewcommand\color[2][]{}%
  }%
  \providecommand\includegraphics[2][]{%
    \GenericError{(gnuplot) \space\space\space\@spaces}{%
      Package graphicx or graphics not loaded%
    }{See the gnuplot documentation for explanation.%
    }{The gnuplot epslatex terminal needs graphicx.sty or graphics.sty.}%
    \renewcommand\includegraphics[2][]{}%
  }%
  \providecommand\rotatebox[2]{#2}%
  \@ifundefined{ifGPcolor}{%
    \newif\ifGPcolor
    \GPcolortrue
  }{}%
  \@ifundefined{ifGPblacktext}{%
    \newif\ifGPblacktext
    \GPblacktexttrue
  }{}%
  \let\gplgaddtomacro\g@addto@macro
  \gdef\gplbacktext{}%
  \gdef\gplfronttext{}%
  \makeatother
  \ifGPblacktext
    \def\colorrgb#1{}%
    \def\colorgray#1{}%
  \else
    \ifGPcolor
      \def\colorrgb#1{\color[rgb]{#1}}%
      \def\colorgray#1{\color[gray]{#1}}%
      \expandafter\def\csname LTw\endcsname{\color{white}}%
      \expandafter\def\csname LTb\endcsname{\color{black}}%
      \expandafter\def\csname LTa\endcsname{\color{black}}%
      \expandafter\def\csname LT0\endcsname{\color[rgb]{1,0,0}}%
      \expandafter\def\csname LT1\endcsname{\color[rgb]{0,1,0}}%
      \expandafter\def\csname LT2\endcsname{\color[rgb]{0,0,1}}%
      \expandafter\def\csname LT3\endcsname{\color[rgb]{1,0,1}}%
      \expandafter\def\csname LT4\endcsname{\color[rgb]{0,1,1}}%
      \expandafter\def\csname LT5\endcsname{\color[rgb]{1,1,0}}%
      \expandafter\def\csname LT6\endcsname{\color[rgb]{0,0,0}}%
      \expandafter\def\csname LT7\endcsname{\color[rgb]{1,0.3,0}}%
      \expandafter\def\csname LT8\endcsname{\color[rgb]{0.5,0.5,0.5}}%
    \else
      \def\colorrgb#1{\color{black}}%
      \def\colorgray#1{\color[gray]{#1}}%
      \expandafter\def\csname LTw\endcsname{\color{white}}%
      \expandafter\def\csname LTb\endcsname{\color{black}}%
      \expandafter\def\csname LTa\endcsname{\color{black}}%
      \expandafter\def\csname LT0\endcsname{\color{black}}%
      \expandafter\def\csname LT1\endcsname{\color{black}}%
      \expandafter\def\csname LT2\endcsname{\color{black}}%
      \expandafter\def\csname LT3\endcsname{\color{black}}%
      \expandafter\def\csname LT4\endcsname{\color{black}}%
      \expandafter\def\csname LT5\endcsname{\color{black}}%
      \expandafter\def\csname LT6\endcsname{\color{black}}%
      \expandafter\def\csname LT7\endcsname{\color{black}}%
      \expandafter\def\csname LT8\endcsname{\color{black}}%
    \fi
  \fi
  \setlength{\unitlength}{0.0500bp}%
  \begin{picture}(4534.00,3400.00)%
    \gplgaddtomacro\gplbacktext{%
      \csname LTb\endcsname%
      \put(384,512){\makebox(0,0)[r]{\strut{}1.0}}%
      \put(384,1086){\makebox(0,0)[r]{\strut{}1.5}}%
      \put(384,1660){\makebox(0,0)[r]{\strut{}2.0}}%
      \put(384,2235){\makebox(0,0)[r]{\strut{}2.5}}%
      \put(384,2809){\makebox(0,0)[r]{\strut{}3.0}}%
      \put(384,3383){\makebox(0,0)[r]{\strut{}3.5}}%
      \put(480,352){\makebox(0,0){\strut{}0.90}}%
      \put(1443,352){\makebox(0,0){\strut{}0.95}}%
      \put(2405,352){\makebox(0,0){\strut{}1.00}}%
      \put(3368,352){\makebox(0,0){\strut{}1.05}}%
      \put(4330,352){\makebox(0,0){\strut{}1.10}}%
      \put(2501,112){\makebox(0,0){\strut{}$\beta$}}%
    }%
    \gplgaddtomacro\gplfronttext{%
      \csname LTb\endcsname%
      \put(3788,3200){\makebox(0,0)[r]{\strut{}$m_{\text{Proton}}$}}%
    }%
    \gplbacktext
    \put(0,0){\includegraphics{MassProton_8_16}}%
    \gplfronttext
  \end{picture}%
\endgroup

%% file: plots/diquarkmass_096.tex
\begingroup
\scriptsize
  \makeatletter
  \providecommand\color[2][]{%
    \GenericError{(gnuplot) \space\space\space\@spaces}{%
      Package color not loaded in conjunction with
      terminal option `colourtext'%
    }{See the gnuplot documentation for explanation.%
    }{Either use 'blacktext' in gnuplot or load the package
      color.sty in LaTeX.}%
    \renewcommand\color[2][]{}%
  }%
  \providecommand\includegraphics[2][]{%
    \GenericError{(gnuplot) \space\space\space\@spaces}{%
      Package graphicx or graphics not loaded%
    }{See the gnuplot documentation for explanation.%
    }{The gnuplot epslatex terminal needs graphicx.sty or graphics.sty.}%
    \renewcommand\includegraphics[2][]{}%
  }%
  \providecommand\rotatebox[2]{#2}%
  \@ifundefined{ifGPcolor}{%
    \newif\ifGPcolor
    \GPcolortrue
  }{}%
  \@ifundefined{ifGPblacktext}{%
    \newif\ifGPblacktext
    \GPblacktexttrue
  }{}%
  \let\gplgaddtomacro\g@addto@macro
  \gdef\gplbacktext{}%
  \gdef\gplfronttext{}%
  \makeatother
  \ifGPblacktext
    \def\colorrgb#1{}%
    \def\colorgray#1{}%
  \else
    \ifGPcolor
      \def\colorrgb#1{\color[rgb]{#1}}%
      \def\colorgray#1{\color[gray]{#1}}%
      \expandafter\def\csname LTw\endcsname{\color{white}}%
      \expandafter\def\csname LTb\endcsname{\color{black}}%
      \expandafter\def\csname LTa\endcsname{\color{black}}%
      \expandafter\def\csname LT0\endcsname{\color[rgb]{1,0,0}}%
      \expandafter\def\csname LT1\endcsname{\color[rgb]{0,1,0}}%
      \expandafter\def\csname LT2\endcsname{\color[rgb]{0,0,1}}%
      \expandafter\def\csname LT3\endcsname{\color[rgb]{1,0,1}}%
      \expandafter\def\csname LT4\endcsname{\color[rgb]{0,1,1}}%
      \expandafter\def\csname LT5\endcsname{\color[rgb]{1,1,0}}%
      \expandafter\def\csname LT6\endcsname{\color[rgb]{0,0,0}}%
      \expandafter\def\csname LT7\endcsname{\color[rgb]{1,0.3,0}}%
      \expandafter\def\csname LT8\endcsname{\color[rgb]{0.5,0.5,0.5}}%
    \else
      \def\colorrgb#1{\color{black}}%
      \def\colorgray#1{\color[gray]{#1}}%
      \expandafter\def\csname LTw\endcsname{\color{white}}%
      \expandafter\def\csname LTb\endcsname{\color{black}}%
      \expandafter\def\csname LTa\endcsname{\color{black}}%
      \expandafter\def\csname LT0\endcsname{\color{black}}%
      \expandafter\def\csname LT1\endcsname{\color{black}}%
      \expandafter\def\csname LT2\endcsname{\color{black}}%
      \expandafter\def\csname LT3\endcsname{\color{black}}%
      \expandafter\def\csname LT4\endcsname{\color{black}}%
      \expandafter\def\csname LT5\endcsname{\color{black}}%
      \expandafter\def\csname LT6\endcsname{\color{black}}%
      \expandafter\def\csname LT7\endcsname{\color{black}}%
      \expandafter\def\csname LT8\endcsname{\color{black}}%
    \fi
  \fi
  \setlength{\unitlength}{0.0500bp}%
  \begin{picture}(4534.00,3400.00)%
    \gplgaddtomacro\gplbacktext{%
      \csname LTb\endcsname%
      \put(384,512){\makebox(0,0)[r]{\strut{}0.2}}%
      \put(384,799){\makebox(0,0)[r]{\strut{}0.3}}%
      \put(384,1086){\makebox(0,0)[r]{\strut{}0.4}}%
      \put(384,1373){\makebox(0,0)[r]{\strut{}0.5}}%
      \put(384,1660){\makebox(0,0)[r]{\strut{}0.6}}%
      \put(384,1947){\makebox(0,0)[r]{\strut{}0.7}}%
      \put(384,2235){\makebox(0,0)[r]{\strut{}0.8}}%
      \put(384,2522){\makebox(0,0)[r]{\strut{}0.9}}%
      \put(384,2809){\makebox(0,0)[r]{\strut{}1.0}}%
      \put(384,3096){\makebox(0,0)[r]{\strut{}1.1}}%
      \put(384,3383){\makebox(0,0)[r]{\strut{}1.2}}%
      \put(985,352){\makebox(0,0){\strut{}0.148}}%
      \put(1996,352){\makebox(0,0){\strut{}0.152}}%
      \put(3007,352){\makebox(0,0){\strut{}0.156}}%
      \put(4018,352){\makebox(0,0){\strut{}0.160}}%
      \put(2501,112){\makebox(0,0){\strut{}$\kappa$}}%
    }%
    \gplgaddtomacro\gplfronttext{%
      \csname LTb\endcsname%
      \put(3788,3200){\makebox(0,0)[r]{\strut{}$d (0^{+})$}}%
      \csname LTb\endcsname%
      \put(3788,2960){\makebox(0,0)[r]{\strut{}$d (1^{+})$}}%
    }%
    \gplbacktext
    \put(0,0){\includegraphics{diquarkmass_096}}%
    \gplfronttext
  \end{picture}%
\endgroup

%% file: plots/mass_105.tex
\begingroup
\scriptsize
  \makeatletter
  \providecommand\color[2][]{%
    \GenericError{(gnuplot) \space\space\space\@spaces}{%
      Package color not loaded in conjunction with
      terminal option `colourtext'%
    }{See the gnuplot documentation for explanation.%
    }{Either use 'blacktext' in gnuplot or load the package
      color.sty in LaTeX.}%
    \renewcommand\color[2][]{}%
  }%
  \providecommand\includegraphics[2][]{%
    \GenericError{(gnuplot) \space\space\space\@spaces}{%
      Package graphicx or graphics not loaded%
    }{See the gnuplot documentation for explanation.%
    }{The gnuplot epslatex terminal needs graphicx.sty or graphics.sty.}%
    \renewcommand\includegraphics[2][]{}%
  }%
  \providecommand\rotatebox[2]{#2}%
  \@ifundefined{ifGPcolor}{%
    \newif\ifGPcolor
    \GPcolortrue
  }{}%
  \@ifundefined{ifGPblacktext}{%
    \newif\ifGPblacktext
    \GPblacktexttrue
  }{}%
  \let\gplgaddtomacro\g@addto@macro
  \gdef\gplbacktext{}%
  \gdef\gplfronttext{}%
  \makeatother
  \ifGPblacktext
    \def\colorrgb#1{}%
    \def\colorgray#1{}%
  \else
    \ifGPcolor
      \def\colorrgb#1{\color[rgb]{#1}}%
      \def\colorgray#1{\color[gray]{#1}}%
      \expandafter\def\csname LTw\endcsname{\color{white}}%
      \expandafter\def\csname LTb\endcsname{\color{black}}%
      \expandafter\def\csname LTa\endcsname{\color{black}}%
      \expandafter\def\csname LT0\endcsname{\color[rgb]{1,0,0}}%
      \expandafter\def\csname LT1\endcsname{\color[rgb]{0,1,0}}%
      \expandafter\def\csname LT2\endcsname{\color[rgb]{0,0,1}}%
      \expandafter\def\csname LT3\endcsname{\color[rgb]{1,0,1}}%
      \expandafter\def\csname LT4\endcsname{\color[rgb]{0,1,1}}%
      \expandafter\def\csname LT5\endcsname{\color[rgb]{1,1,0}}%
      \expandafter\def\csname LT6\endcsname{\color[rgb]{0,0,0}}%
      \expandafter\def\csname LT7\endcsname{\color[rgb]{1,0.3,0}}%
      \expandafter\def\csname LT8\endcsname{\color[rgb]{0.5,0.5,0.5}}%
    \else
      \def\colorrgb#1{\color{black}}%
      \def\colorgray#1{\color[gray]{#1}}%
      \expandafter\def\csname LTw\endcsname{\color{white}}%
      \expandafter\def\csname LTb\endcsname{\color{black}}%
      \expandafter\def\csname LTa\endcsname{\color{black}}%
      \expandafter\def\csname LT0\endcsname{\color{black}}%
      \expandafter\def\csname LT1\endcsname{\color{black}}%
      \expandafter\def\csname LT2\endcsname{\color{black}}%
      \expandafter\def\csname LT3\endcsname{\color{black}}%
      \expandafter\def\csname LT4\endcsname{\color{black}}%
      \expandafter\def\csname LT5\endcsname{\color{black}}%
      \expandafter\def\csname LT6\endcsname{\color{black}}%
      \expandafter\def\csname LT7\endcsname{\color{black}}%
      \expandafter\def\csname LT8\endcsname{\color{black}}%
    \fi
  \fi
  \setlength{\unitlength}{0.0500bp}%
  \begin{picture}(4534.00,3400.00)%
    \gplgaddtomacro\gplbacktext{%
      \csname LTb\endcsname%
      \put(384,320){\makebox(0,0)[r]{\strut{}0.0}}%
      \put(384,723){\makebox(0,0)[r]{\strut{}0.5}}%
      \put(384,1126){\makebox(0,0)[r]{\strut{}1.0}}%
      \put(384,1529){\makebox(0,0)[r]{\strut{}1.5}}%
      \put(384,1932){\makebox(0,0)[r]{\strut{}2.0}}%
      \put(384,2335){\makebox(0,0)[r]{\strut{}2.5}}%
      \put(384,2738){\makebox(0,0)[r]{\strut{}3.0}}%
      \put(384,3141){\makebox(0,0)[r]{\strut{}3.5}}%
      \put(480,160){\makebox(0,0){\strut{}}}%
      \put(929,160){\makebox(0,0){\strut{}$0^+$}}%
      \put(1378,160){\makebox(0,0){\strut{}$0^-$}}%
      \put(1828,160){\makebox(0,0){\strut{}$\frac{1}{2}^+$}}%
      \put(2277,160){\makebox(0,0){\strut{}$\frac{1}{2}^-$}}%
      \put(2726,160){\makebox(0,0){\strut{}$1^+$}}%
      \put(3175,160){\makebox(0,0){\strut{}$1^-$}}%
      \put(3625,160){\makebox(0,0){\strut{}$\frac{3}{2}^+$}}%
      \put(4074,160){\makebox(0,0){\strut{}$\frac{3}{2}^-$}}%
      \put(4523,160){\makebox(0,0){\strut{}}}%
      \put(4619,320){\makebox(0,0)[l]{\strut{}0}}%
      \put(4619,796){\makebox(0,0)[l]{\strut{}326}}%
      \put(4619,1050){\makebox(0,0)[l]{\strut{}500}}%
      \put(4619,1690){\makebox(0,0)[l]{\strut{}938}}%
      \put(4619,2511){\makebox(0,0)[l]{\strut{}1500}}%
      \put(4619,3241){\makebox(0,0)[l]{\strut{}2000}}%
      \put(208,1771){\makebox(0,0){\strut{}$m$}}%
      \put(4890,1371){\makebox(0,0){\strut{}$m$ in MeV}}%
    }%
    \gplgaddtomacro\gplfronttext{%
      \csname LTb\endcsname%
      \put(929,796){\makebox(0,0){\strut{}$d$}}%
      \put(1289,779){\makebox(0,0){\strut{}$d$}}%
      \put(2726,844){\makebox(0,0){\strut{}$d$}}%
      \put(3175,812){\makebox(0,0){\strut{}$d$}}%
      \put(1558,820){\makebox(0,0){\strut{}$\eta$}}%
      \put(1828,1690){\makebox(0,0){\strut{}$N$}}%
      \put(2277,1779){\makebox(0,0){\strut{}$N$}}%
      \put(3625,1723){\makebox(0,0){\strut{}$\Delta$}}%
      \put(4074,1819){\makebox(0,0){\strut{}$\Delta$}}%
      \put(929,1771){\makebox(0,0){\strut{}$d^*$}}%
      \put(1289,1811){\makebox(0,0){\strut{}$d^*$}}%
      \put(2726,1876){\makebox(0,0){\strut{}$d^*$}}%
      \put(3175,1795){\makebox(0,0){\strut{}$d^*$}}%
      \put(1558,1835){\makebox(0,0){\strut{}$\eta^*$}}%
      \put(1828,3036){\makebox(0,0){\strut{}$N^*$}}%
      \put(2277,3101){\makebox(0,0){\strut{}$N^*$}}%
      \put(3625,3109){\makebox(0,0){\strut{}$\Delta^*$}}%
      \put(4074,3117){\makebox(0,0){\strut{}$\Delta^*$}}%
    }%
    \gplbacktext
    \put(0,0){\includegraphics{mass_105}}%
    \gplfronttext
  \end{picture}%
\endgroup

%% file: plots/mass_096.tex
\begingroup
\scriptsize
  \makeatletter
  \providecommand\color[2][]{%
    \GenericError{(gnuplot) \space\space\space\@spaces}{%
      Package color not loaded in conjunction with
      terminal option `colourtext'%
    }{See the gnuplot documentation for explanation.%
    }{Either use 'blacktext' in gnuplot or load the package
      color.sty in LaTeX.}%
    \renewcommand\color[2][]{}%
  }%
  \providecommand\includegraphics[2][]{%
    \GenericError{(gnuplot) \space\space\space\@spaces}{%
      Package graphicx or graphics not loaded%
    }{See the gnuplot documentation for explanation.%
    }{The gnuplot epslatex terminal needs graphicx.sty or graphics.sty.}%
    \renewcommand\includegraphics[2][]{}%
  }%
  \providecommand\rotatebox[2]{#2}%
  \@ifundefined{ifGPcolor}{%
    \newif\ifGPcolor
    \GPcolortrue
  }{}%
  \@ifundefined{ifGPblacktext}{%
    \newif\ifGPblacktext
    \GPblacktexttrue
  }{}%
  \let\gplgaddtomacro\g@addto@macro
  \gdef\gplbacktext{}%
  \gdef\gplfronttext{}%
  \makeatother
  \ifGPblacktext
    \def\colorrgb#1{}%
    \def\colorgray#1{}%
  \else
    \ifGPcolor
      \def\colorrgb#1{\color[rgb]{#1}}%
      \def\colorgray#1{\color[gray]{#1}}%
      \expandafter\def\csname LTw\endcsname{\color{white}}%
      \expandafter\def\csname LTb\endcsname{\color{black}}%
      \expandafter\def\csname LTa\endcsname{\color{black}}%
      \expandafter\def\csname LT0\endcsname{\color[rgb]{1,0,0}}%
      \expandafter\def\csname LT1\endcsname{\color[rgb]{0,1,0}}%
      \expandafter\def\csname LT2\endcsname{\color[rgb]{0,0,1}}%
      \expandafter\def\csname LT3\endcsname{\color[rgb]{1,0,1}}%
      \expandafter\def\csname LT4\endcsname{\color[rgb]{0,1,1}}%
      \expandafter\def\csname LT5\endcsname{\color[rgb]{1,1,0}}%
      \expandafter\def\csname LT6\endcsname{\color[rgb]{0,0,0}}%
      \expandafter\def\csname LT7\endcsname{\color[rgb]{1,0.3,0}}%
      \expandafter\def\csname LT8\endcsname{\color[rgb]{0.5,0.5,0.5}}%
    \else
      \def\colorrgb#1{\color{black}}%
      \def\colorgray#1{\color[gray]{#1}}%
      \expandafter\def\csname LTw\endcsname{\color{white}}%
      \expandafter\def\csname LTb\endcsname{\color{black}}%
      \expandafter\def\csname LTa\endcsname{\color{black}}%
      \expandafter\def\csname LT0\endcsname{\color{black}}%
      \expandafter\def\csname LT1\endcsname{\color{black}}%
      \expandafter\def\csname LT2\endcsname{\color{black}}%
      \expandafter\def\csname LT3\endcsname{\color{black}}%
      \expandafter\def\csname LT4\endcsname{\color{black}}%
      \expandafter\def\csname LT5\endcsname{\color{black}}%
      \expandafter\def\csname LT6\endcsname{\color{black}}%
      \expandafter\def\csname LT7\endcsname{\color{black}}%
      \expandafter\def\csname LT8\endcsname{\color{black}}%
    \fi
  \fi
  \setlength{\unitlength}{0.0500bp}%
  \begin{picture}(4534.00,3400.00)%
    \gplgaddtomacro\gplbacktext{%
      \csname LTb\endcsname%
      \put(384,320){\makebox(0,0)[r]{\strut{}0.0}}%
      \put(384,723){\makebox(0,0)[r]{\strut{}0.5}}%
      \put(384,1126){\makebox(0,0)[r]{\strut{}1.0}}%
      \put(384,1529){\makebox(0,0)[r]{\strut{}1.5}}%
      \put(384,1932){\makebox(0,0)[r]{\strut{}2.0}}%
      \put(384,2335){\makebox(0,0)[r]{\strut{}2.5}}%
      \put(384,2738){\makebox(0,0)[r]{\strut{}3.0}}%
      \put(384,3141){\makebox(0,0)[r]{\strut{}3.5}}%
      \put(480,160){\makebox(0,0){\strut{}}}%
      \put(929,160){\makebox(0,0){\strut{}$0^+$}}%
      \put(1378,160){\makebox(0,0){\strut{}$0^-$}}%
      \put(1828,160){\makebox(0,0){\strut{}$\frac{1}{2}^+$}}%
      \put(2277,160){\makebox(0,0){\strut{}$\frac{1}{2}^-$}}%
      \put(2726,160){\makebox(0,0){\strut{}$1^+$}}%
      \put(3175,160){\makebox(0,0){\strut{}$1^-$}}%
      \put(3625,160){\makebox(0,0){\strut{}$\frac{3}{2}^+$}}%
      \put(4074,160){\makebox(0,0){\strut{}$\frac{3}{2}^-$}}%
      \put(4523,160){\makebox(0,0){\strut{}}}%
      \put(4619,320){\makebox(0,0)[l]{\strut{}0}}%
      \put(4619,666){\makebox(0,0)[l]{\strut{}247}}%
      \put(4619,1020){\makebox(0,0)[l]{\strut{}500}}%
      \put(4619,1634){\makebox(0,0)[l]{\strut{}938}}%
      \put(4619,2421){\makebox(0,0)[l]{\strut{}1500}}%
      \put(4619,3121){\makebox(0,0)[l]{\strut{}2000}}%
      \put(208,1771){\makebox(0,0){\strut{}$m$}}%
      \put(4890,1371){\makebox(0,0){\strut{}$m$ in MeV}}%
    }%
    \gplgaddtomacro\gplfronttext{%
      \csname LTb\endcsname%
      \put(929,667){\makebox(0,0){\strut{}$d$}}%
      \put(1378,836){\makebox(0,0){\strut{}$d$}}%
      \put(2726,852){\makebox(0,0){\strut{}$d$}}%
      \put(3175,1231){\makebox(0,0){\strut{}$d$}}%
      \put(1378,715){\makebox(0,0){\strut{}$\eta$}}%
      \put(1828,1634){\makebox(0,0){\strut{}$N$}}%
      \put(2277,1747){\makebox(0,0){\strut{}$N$}}%
      \put(3625,1763){\makebox(0,0){\strut{}$\Delta$}}%
      \put(4074,1876){\makebox(0,0){\strut{}$\Delta$}}%
      \put(929,1666){\makebox(0,0){\strut{}$d^*$}}%
      \put(1378,1884){\makebox(0,0){\strut{}$d^*$}}%
      \put(2726,1835){\makebox(0,0){\strut{}$d^*$}}%
      \put(3175,2271){\makebox(0,0){\strut{}$d^*$}}%
      \put(1378,1714){\makebox(0,0){\strut{}$\eta^*$}}%
      \put(1828,2932){\makebox(0,0){\strut{}$N^*$}}%
      \put(2277,3012){\makebox(0,0){\strut{}$N^*$}}%
      \put(3625,3020){\makebox(0,0){\strut{}$\Delta^*$}}%
      \put(4074,3028){\makebox(0,0){\strut{}$\Delta^*$}}%
    }%
    \gplbacktext
    \put(0,0){\includegraphics{mass_096}}%
    \gplfronttext
  \end{picture}%
\endgroup

%% file: plots/pd8_16_QND_105.tex
\begingroup
\scriptsize
  \makeatletter
  \providecommand\color[2][]{%
    \GenericError{(gnuplot) \space\space\space\@spaces}{%
      Package color not loaded in conjunction with
      terminal option `colourtext'%
    }{See the gnuplot documentation for explanation.%
    }{Either use 'blacktext' in gnuplot or load the package
      color.sty in LaTeX.}%
    \renewcommand\color[2][]{}%
  }%
  \providecommand\includegraphics[2][]{%
    \GenericError{(gnuplot) \space\space\space\@spaces}{%
      Package graphicx or graphics not loaded%
    }{See the gnuplot documentation for explanation.%
    }{The gnuplot epslatex terminal needs graphicx.sty or graphics.sty.}%
    \renewcommand\includegraphics[2][]{}%
  }%
  \providecommand\rotatebox[2]{#2}%
  \@ifundefined{ifGPcolor}{%
    \newif\ifGPcolor
    \GPcolortrue
  }{}%
  \@ifundefined{ifGPblacktext}{%
    \newif\ifGPblacktext
    \GPblacktexttrue
  }{}%
  \let\gplgaddtomacro\g@addto@macro
  \gdef\gplbacktext{}%
  \gdef\gplfronttext{}%
  \makeatother
  \ifGPblacktext
    \def\colorrgb#1{}%
    \def\colorgray#1{}%
  \else
    \ifGPcolor
      \def\colorrgb#1{\color[rgb]{#1}}%
      \def\colorgray#1{\color[gray]{#1}}%
      \expandafter\def\csname LTw\endcsname{\color{white}}%
      \expandafter\def\csname LTb\endcsname{\color{black}}%
      \expandafter\def\csname LTa\endcsname{\color{black}}%
      \expandafter\def\csname LT0\endcsname{\color[rgb]{1,0,0}}%
      \expandafter\def\csname LT1\endcsname{\color[rgb]{0,1,0}}%
      \expandafter\def\csname LT2\endcsname{\color[rgb]{0,0,1}}%
      \expandafter\def\csname LT3\endcsname{\color[rgb]{1,0,1}}%
      \expandafter\def\csname LT4\endcsname{\color[rgb]{0,1,1}}%
      \expandafter\def\csname LT5\endcsname{\color[rgb]{1,1,0}}%
      \expandafter\def\csname LT6\endcsname{\color[rgb]{0,0,0}}%
      \expandafter\def\csname LT7\endcsname{\color[rgb]{1,0.3,0}}%
      \expandafter\def\csname LT8\endcsname{\color[rgb]{0.5,0.5,0.5}}%
    \else
      \def\colorrgb#1{\color{black}}%
      \def\colorgray#1{\color[gray]{#1}}%
      \expandafter\def\csname LTw\endcsname{\color{white}}%
      \expandafter\def\csname LTb\endcsname{\color{black}}%
      \expandafter\def\csname LTa\endcsname{\color{black}}%
      \expandafter\def\csname LT0\endcsname{\color{black}}%
      \expandafter\def\csname LT1\endcsname{\color{black}}%
      \expandafter\def\csname LT2\endcsname{\color{black}}%
      \expandafter\def\csname LT3\endcsname{\color{black}}%
      \expandafter\def\csname LT4\endcsname{\color{black}}%
      \expandafter\def\csname LT5\endcsname{\color{black}}%
      \expandafter\def\csname LT6\endcsname{\color{black}}%
      \expandafter\def\csname LT7\endcsname{\color{black}}%
      \expandafter\def\csname LT8\endcsname{\color{black}}%
    \fi
  \fi
  \setlength{\unitlength}{0.0500bp}%
  \begin{picture}(4534.00,3400.00)%
    \gplgaddtomacro\gplbacktext{%
      \csname LTb\endcsname%
      \put(384,531){\makebox(0,0)[r]{\strut{}0}}%
      \put(384,911){\makebox(0,0)[r]{\strut{}2}}%
      \put(384,1292){\makebox(0,0)[r]{\strut{}4}}%
      \put(384,1672){\makebox(0,0)[r]{\strut{}6}}%
      \put(384,2052){\makebox(0,0)[r]{\strut{}8}}%
      \put(384,2432){\makebox(0,0)[r]{\strut{}10}}%
      \put(384,2813){\makebox(0,0)[r]{\strut{}12}}%
      \put(384,3193){\makebox(0,0)[r]{\strut{}14}}%
      \put(480,352){\makebox(0,0){\strut{}0.0}}%
      \put(884,352){\makebox(0,0){\strut{}0.2}}%
      \put(1289,352){\makebox(0,0){\strut{}0.4}}%
      \put(1693,352){\makebox(0,0){\strut{}0.6}}%
      \put(2097,352){\makebox(0,0){\strut{}0.8}}%
      \put(2502,352){\makebox(0,0){\strut{}1.0}}%
      \put(2906,352){\makebox(0,0){\strut{}1.2}}%
      \put(3310,352){\makebox(0,0){\strut{}1.4}}%
      \put(3714,352){\makebox(0,0){\strut{}1.6}}%
      \put(4119,352){\makebox(0,0){\strut{}1.8}}%
      \put(4523,352){\makebox(0,0){\strut{}2.0}}%
      \put(208,2187){\makebox(0,0){\strut{}$n_q$}}%
      \put(2501,112){\makebox(0,0){\strut{}$\mu$}}%
    }%
    \gplgaddtomacro\gplfronttext{%
    }%
    \gplbacktext
    \put(0,0){\includegraphics{pd8_16_QND_105}}%
    \gplfronttext
  \end{picture}%
\endgroup

%% file: plots/onset.tex
\begingroup
\scriptsize
  \makeatletter
  \providecommand\color[2][]{%
    \GenericError{(gnuplot) \space\space\space\@spaces}{%
      Package color not loaded in conjunction with
      terminal option `colourtext'%
    }{See the gnuplot documentation for explanation.%
    }{Either use 'blacktext' in gnuplot or load the package
      color.sty in LaTeX.}%
    \renewcommand\color[2][]{}%
  }%
  \providecommand\includegraphics[2][]{%
    \GenericError{(gnuplot) \space\space\space\@spaces}{%
      Package graphicx or graphics not loaded%
    }{See the gnuplot documentation for explanation.%
    }{The gnuplot epslatex terminal needs graphicx.sty or graphics.sty.}%
    \renewcommand\includegraphics[2][]{}%
  }%
  \providecommand\rotatebox[2]{#2}%
  \@ifundefined{ifGPcolor}{%
    \newif\ifGPcolor
    \GPcolortrue
  }{}%
  \@ifundefined{ifGPblacktext}{%
    \newif\ifGPblacktext
    \GPblacktexttrue
  }{}%
  \let\gplgaddtomacro\g@addto@macro
  \gdef\gplbacktext{}%
  \gdef\gplfronttext{}%
  \makeatother
  \ifGPblacktext
    \def\colorrgb#1{}%
    \def\colorgray#1{}%
  \else
    \ifGPcolor
      \def\colorrgb#1{\color[rgb]{#1}}%
      \def\colorgray#1{\color[gray]{#1}}%
      \expandafter\def\csname LTw\endcsname{\color{white}}%
      \expandafter\def\csname LTb\endcsname{\color{black}}%
      \expandafter\def\csname LTa\endcsname{\color{black}}%
      \expandafter\def\csname LT0\endcsname{\color[rgb]{1,0,0}}%
      \expandafter\def\csname LT1\endcsname{\color[rgb]{0,1,0}}%
      \expandafter\def\csname LT2\endcsname{\color[rgb]{0,0,1}}%
      \expandafter\def\csname LT3\endcsname{\color[rgb]{1,0,1}}%
      \expandafter\def\csname LT4\endcsname{\color[rgb]{0,1,1}}%
      \expandafter\def\csname LT5\endcsname{\color[rgb]{1,1,0}}%
      \expandafter\def\csname LT6\endcsname{\color[rgb]{0,0,0}}%
      \expandafter\def\csname LT7\endcsname{\color[rgb]{1,0.3,0}}%
      \expandafter\def\csname LT8\endcsname{\color[rgb]{0.5,0.5,0.5}}%
    \else
      \def\colorrgb#1{\color{black}}%
      \def\colorgray#1{\color[gray]{#1}}%
      \expandafter\def\csname LTw\endcsname{\color{white}}%
      \expandafter\def\csname LTb\endcsname{\color{black}}%
      \expandafter\def\csname LTa\endcsname{\color{black}}%
      \expandafter\def\csname LT0\endcsname{\color{black}}%
      \expandafter\def\csname LT1\endcsname{\color{black}}%
      \expandafter\def\csname LT2\endcsname{\color{black}}%
      \expandafter\def\csname LT3\endcsname{\color{black}}%
      \expandafter\def\csname LT4\endcsname{\color{black}}%
      \expandafter\def\csname LT5\endcsname{\color{black}}%
      \expandafter\def\csname LT6\endcsname{\color{black}}%
      \expandafter\def\csname LT7\endcsname{\color{black}}%
      \expandafter\def\csname LT8\endcsname{\color{black}}%
    \fi
  \fi
  \setlength{\unitlength}{0.0500bp}%
  \begin{picture}(4534.00,3400.00)%
    \gplgaddtomacro\gplbacktext{%
      \csname LTb\endcsname%
      \put(384,512){\makebox(0,0)[r]{\strut{}0.15}}%
      \put(384,831){\makebox(0,0)[r]{\strut{}0.20}}%
      \put(384,1150){\makebox(0,0)[r]{\strut{}0.25}}%
      \put(384,1469){\makebox(0,0)[r]{\strut{}0.30}}%
      \put(384,1788){\makebox(0,0)[r]{\strut{}0.35}}%
      \put(384,2107){\makebox(0,0)[r]{\strut{}0.40}}%
      \put(384,2426){\makebox(0,0)[r]{\strut{}0.45}}%
      \put(384,2745){\makebox(0,0)[r]{\strut{}0.50}}%
      \put(384,3064){\makebox(0,0)[r]{\strut{}0.55}}%
      \put(384,3383){\makebox(0,0)[r]{\strut{}0.60}}%
      \put(733,352){\makebox(0,0){\strut{}0.96}}%
      \put(1238,352){\makebox(0,0){\strut{}0.98}}%
      \put(1743,352){\makebox(0,0){\strut{}1.00}}%
      \put(2249,352){\makebox(0,0){\strut{}1.02}}%
      \put(2754,352){\makebox(0,0){\strut{}1.04}}%
      \put(3260,352){\makebox(0,0){\strut{}1.06}}%
      \put(3765,352){\makebox(0,0){\strut{}1.08}}%
      \put(4270,352){\makebox(0,0){\strut{}1.10}}%
      \put(-176,1947){\makebox(0,0){\strut{}$\mu$}}%
      \put(2501,112){\makebox(0,0){\strut{}$\beta$}}%
    }%
    \gplgaddtomacro\gplfronttext{%
      \csname LTb\endcsname%
      \put(3788,3200){\makebox(0,0)[r]{\strut{}$m_{d(0^+)}/2$}}%
    }%
    \gplbacktext
    \put(0,0){\includegraphics{onset}}%
    \gplfronttext
  \end{picture}%
\endgroup

%% file: plots/qnd_105.tex
\begingroup
\scriptsize
  \makeatletter
  \providecommand\color[2][]{%
    \GenericError{(gnuplot) \space\space\space\@spaces}{%
      Package color not loaded in conjunction with
      terminal option `colourtext'%
    }{See the gnuplot documentation for explanation.%
    }{Either use 'blacktext' in gnuplot or load the package
      color.sty in LaTeX.}%
    \renewcommand\color[2][]{}%
  }%
  \providecommand\includegraphics[2][]{%
    \GenericError{(gnuplot) \space\space\space\@spaces}{%
      Package graphicx or graphics not loaded%
    }{See the gnuplot documentation for explanation.%
    }{The gnuplot epslatex terminal needs graphicx.sty or graphics.sty.}%
    \renewcommand\includegraphics[2][]{}%
  }%
  \providecommand\rotatebox[2]{#2}%
  \@ifundefined{ifGPcolor}{%
    \newif\ifGPcolor
    \GPcolortrue
  }{}%
  \@ifundefined{ifGPblacktext}{%
    \newif\ifGPblacktext
    \GPblacktexttrue
  }{}%
  \let\gplgaddtomacro\g@addto@macro
  \gdef\gplbacktext{}%
  \gdef\gplfronttext{}%
  \makeatother
  \ifGPblacktext
    \def\colorrgb#1{}%
    \def\colorgray#1{}%
  \else
    \ifGPcolor
      \def\colorrgb#1{\color[rgb]{#1}}%
      \def\colorgray#1{\color[gray]{#1}}%
      \expandafter\def\csname LTw\endcsname{\color{white}}%
      \expandafter\def\csname LTb\endcsname{\color{black}}%
      \expandafter\def\csname LTa\endcsname{\color{black}}%
      \expandafter\def\csname LT0\endcsname{\color[rgb]{1,0,0}}%
      \expandafter\def\csname LT1\endcsname{\color[rgb]{0,1,0}}%
      \expandafter\def\csname LT2\endcsname{\color[rgb]{0,0,1}}%
      \expandafter\def\csname LT3\endcsname{\color[rgb]{1,0,1}}%
      \expandafter\def\csname LT4\endcsname{\color[rgb]{0,1,1}}%
      \expandafter\def\csname LT5\endcsname{\color[rgb]{1,1,0}}%
      \expandafter\def\csname LT6\endcsname{\color[rgb]{0,0,0}}%
      \expandafter\def\csname LT7\endcsname{\color[rgb]{1,0.3,0}}%
      \expandafter\def\csname LT8\endcsname{\color[rgb]{0.5,0.5,0.5}}%
    \else
      \def\colorrgb#1{\color{black}}%
      \def\colorgray#1{\color[gray]{#1}}%
      \expandafter\def\csname LTw\endcsname{\color{white}}%
      \expandafter\def\csname LTb\endcsname{\color{black}}%
      \expandafter\def\csname LTa\endcsname{\color{black}}%
      \expandafter\def\csname LT0\endcsname{\color{black}}%
      \expandafter\def\csname LT1\endcsname{\color{black}}%
      \expandafter\def\csname LT2\endcsname{\color{black}}%
      \expandafter\def\csname LT3\endcsname{\color{black}}%
      \expandafter\def\csname LT4\endcsname{\color{black}}%
      \expandafter\def\csname LT5\endcsname{\color{black}}%
      \expandafter\def\csname LT6\endcsname{\color{black}}%
      \expandafter\def\csname LT7\endcsname{\color{black}}%
      \expandafter\def\csname LT8\endcsname{\color{black}}%
    \fi
  \fi
  \setlength{\unitlength}{0.0500bp}%
  \begin{picture}(4534.00,3400.00)%
    \gplgaddtomacro\gplbacktext{%
      \csname LTb\endcsname%
      \put(384,717){\makebox(0,0)[r]{\strut{}0.00}}%
      \put(384,1127){\makebox(0,0)[r]{\strut{}0.02}}%
      \put(384,1537){\makebox(0,0)[r]{\strut{}0.04}}%
      \put(384,1948){\makebox(0,0)[r]{\strut{}0.06}}%
      \put(384,2358){\makebox(0,0)[r]{\strut{}0.08}}%
      \put(384,2768){\makebox(0,0)[r]{\strut{}0.10}}%
      \put(384,3178){\makebox(0,0)[r]{\strut{}0.12}}%
      \put(480,352){\makebox(0,0){\strut{}0.00}}%
      \put(985,352){\makebox(0,0){\strut{}0.10}}%
      \put(1491,352){\makebox(0,0){\strut{}0.20}}%
      \put(1996,352){\makebox(0,0){\strut{}0.30}}%
      \put(2502,352){\makebox(0,0){\strut{}0.40}}%
      \put(3007,352){\makebox(0,0){\strut{}0.50}}%
      \put(3512,352){\makebox(0,0){\strut{}0.60}}%
      \put(4018,352){\makebox(0,0){\strut{}0.70}}%
      \put(4523,352){\makebox(0,0){\strut{}0.80}}%
      \put(-176,1947){\makebox(0,0){\strut{}$n_q$}}%
      \put(2501,112){\makebox(0,0){\strut{}$\mu$}}%
    }%
    \gplgaddtomacro\gplfronttext{%
      \csname LTb\endcsname%
      \put(1971,1537){\makebox(0,0){\strut{}$d(0^{+})$}}%
      \put(1920,1947){\makebox(0,0){\strut{}$d(0^{-})$}}%
      \put(3411,2358){\makebox(0,0){\strut{}$\Delta(\frac{3}{2}^{+})$}}%
      \put(3613,2768){\makebox(0,0){\strut{}$\Delta(\frac{3}{2}^{-})$}}%
    }%
    \gplbacktext
    \put(0,0){\includegraphics{qnd_105}}%
    \gplfronttext
  \end{picture}%
\endgroup

%% file: plots/qnd_096.tex
\begingroup
\scriptsize
  \makeatletter
  \providecommand\color[2][]{%
    \GenericError{(gnuplot) \space\space\space\@spaces}{%
      Package color not loaded in conjunction with
      terminal option `colourtext'%
    }{See the gnuplot documentation for explanation.%
    }{Either use 'blacktext' in gnuplot or load the package
      color.sty in LaTeX.}%
    \renewcommand\color[2][]{}%
  }%
  \providecommand\includegraphics[2][]{%
    \GenericError{(gnuplot) \space\space\space\@spaces}{%
      Package graphicx or graphics not loaded%
    }{See the gnuplot documentation for explanation.%
    }{The gnuplot epslatex terminal needs graphicx.sty or graphics.sty.}%
    \renewcommand\includegraphics[2][]{}%
  }%
  \providecommand\rotatebox[2]{#2}%
  \@ifundefined{ifGPcolor}{%
    \newif\ifGPcolor
    \GPcolortrue
  }{}%
  \@ifundefined{ifGPblacktext}{%
    \newif\ifGPblacktext
    \GPblacktexttrue
  }{}%
  \let\gplgaddtomacro\g@addto@macro
  \gdef\gplbacktext{}%
  \gdef\gplfronttext{}%
  \makeatother
  \ifGPblacktext
    \def\colorrgb#1{}%
    \def\colorgray#1{}%
  \else
    \ifGPcolor
      \def\colorrgb#1{\color[rgb]{#1}}%
      \def\colorgray#1{\color[gray]{#1}}%
      \expandafter\def\csname LTw\endcsname{\color{white}}%
      \expandafter\def\csname LTb\endcsname{\color{black}}%
      \expandafter\def\csname LTa\endcsname{\color{black}}%
      \expandafter\def\csname LT0\endcsname{\color[rgb]{1,0,0}}%
      \expandafter\def\csname LT1\endcsname{\color[rgb]{0,1,0}}%
      \expandafter\def\csname LT2\endcsname{\color[rgb]{0,0,1}}%
      \expandafter\def\csname LT3\endcsname{\color[rgb]{1,0,1}}%
      \expandafter\def\csname LT4\endcsname{\color[rgb]{0,1,1}}%
      \expandafter\def\csname LT5\endcsname{\color[rgb]{1,1,0}}%
      \expandafter\def\csname LT6\endcsname{\color[rgb]{0,0,0}}%
      \expandafter\def\csname LT7\endcsname{\color[rgb]{1,0.3,0}}%
      \expandafter\def\csname LT8\endcsname{\color[rgb]{0.5,0.5,0.5}}%
    \else
      \def\colorrgb#1{\color{black}}%
      \def\colorgray#1{\color[gray]{#1}}%
      \expandafter\def\csname LTw\endcsname{\color{white}}%
      \expandafter\def\csname LTb\endcsname{\color{black}}%
      \expandafter\def\csname LTa\endcsname{\color{black}}%
      \expandafter\def\csname LT0\endcsname{\color{black}}%
      \expandafter\def\csname LT1\endcsname{\color{black}}%
      \expandafter\def\csname LT2\endcsname{\color{black}}%
      \expandafter\def\csname LT3\endcsname{\color{black}}%
      \expandafter\def\csname LT4\endcsname{\color{black}}%
      \expandafter\def\csname LT5\endcsname{\color{black}}%
      \expandafter\def\csname LT6\endcsname{\color{black}}%
      \expandafter\def\csname LT7\endcsname{\color{black}}%
      \expandafter\def\csname LT8\endcsname{\color{black}}%
    \fi
  \fi
  \setlength{\unitlength}{0.0500bp}%
  \begin{picture}(4534.00,3400.00)%
    \gplgaddtomacro\gplbacktext{%
      \csname LTb\endcsname%
      \put(384,717){\makebox(0,0)[r]{\strut{}0.00}}%
      \put(384,1127){\makebox(0,0)[r]{\strut{}0.02}}%
      \put(384,1537){\makebox(0,0)[r]{\strut{}0.04}}%
      \put(384,1948){\makebox(0,0)[r]{\strut{}0.06}}%
      \put(384,2358){\makebox(0,0)[r]{\strut{}0.08}}%
      \put(384,2768){\makebox(0,0)[r]{\strut{}0.10}}%
      \put(384,3178){\makebox(0,0)[r]{\strut{}0.12}}%
      \put(480,352){\makebox(0,0){\strut{}0.00}}%
      \put(985,352){\makebox(0,0){\strut{}0.10}}%
      \put(1491,352){\makebox(0,0){\strut{}0.20}}%
      \put(1996,352){\makebox(0,0){\strut{}0.30}}%
      \put(2502,352){\makebox(0,0){\strut{}0.40}}%
      \put(3007,352){\makebox(0,0){\strut{}0.50}}%
      \put(3512,352){\makebox(0,0){\strut{}0.60}}%
      \put(4018,352){\makebox(0,0){\strut{}0.70}}%
      \put(4523,352){\makebox(0,0){\strut{}0.80}}%
      \put(-176,1947){\makebox(0,0){\strut{}$n_q$}}%
      \put(2501,112){\makebox(0,0){\strut{}$\mu$}}%
    }%
    \gplgaddtomacro\gplfronttext{%
      \csname LTb\endcsname%
      \put(1567,1127){\makebox(0,0){\strut{}$d(0^{+})$}}%
      \put(2097,1537){\makebox(0,0){\strut{}$d(0^{-})$}}%
      \put(3495,2563){\makebox(0,0){\strut{}$\Delta(\frac{3}{2}^{+})$}}%
      \put(3731,2973){\makebox(0,0){\strut{}$\Delta(\frac{3}{2}^{-})$}}%
    }%
    \gplbacktext
    \put(0,0){\includegraphics{qnd_096}}%
    \gplfronttext
  \end{picture}%
\endgroup

%% file: paper.bbl
\begin{thebibliography}{99}

\providecommand{\eprint}[1]{ [\href{http://arxiv.org/abs/#1}{arXiv:#1}]}

\bibitem{Gattringer:2010zz}
C.~Gattringer and C.B.~Lang,
\newblock Lect.Notes Phys. 788 (2010) 1--343


\bibitem{Kogut:2000ek}
  J.B.~Kogut, M.A.~Stephanov et al.,
 \newblock{Nucl.Phys. B 582 (2000) 477--513 

\bibitem{Hands:2011ye}
      S.~Hands, P.~Kenny, S.~Kim and J.~Skullerud,
      \newblock Eur.Phys.J. A 47 (2011) 60

\bibitem{Holland:2003jy}
  K.~Holland, P.~Minkowski, M.~Pepe and U.J.~Wiese},
 \newblock Nucl. Phys. B 668 (2003) 207--236
      
\bibitem{Maas:2012wr}
A.~Maas, L.v.~Smekal, B.H.~Wellegehausen and A.~Wipf,
\newblock Phys.Rev. D 86 (2012)
   
\bibitem{Hands:2000ei}
  S.~Hands, I.~Montvay, S.~Morrison, M.~Oevers and L.~Scorzato,
  \newblock Eur.Phys.J. C 17 (2000) 285--302

 


\end{thebibliography}
